# A Method for Classifying Orbits near Asteroids

Xianyu Wang, Shengping Gong, and Junfeng Li

*Department of Aerospace Engineering, Tsinghua University, Beijing 100084 CHINA*

*Email: wangxianyu12@mails.tsinghua.edu.cn; gsp04@mails.tsinghua.edu.cn;*

*lijunf@mail.tsinghua.edu.cn*

**Abstract** A method for classifying orbits near asteroids under a polyhedral gravitational field is presented, and may serve as a valuable reference for spacecraft orbit design for asteroid exploration. The orbital dynamics near asteroids are very complex. According to the variation in orbit characteristics after being affected by gravitational perturbation during the periapsis passage, orbits near an asteroid can be classified into 9 categories: "surrounding-to-surrounding", "surrounding-to-surface", "surrounding-to-infinity", "infinity-to-infinity", "infinity-to-surface", "infinity-to-surrounding", "surface-to-surface", "surface-to-surrounding", and "surface-to-infinity". Assume that the orbital elements are constant near the periapsis, the gravitation potential is expanded into a harmonic series. Then, the influence of the gravitational perturbation on the orbit is studied analytically. The styles of orbits are dependent on the argument of periapsis, the periapsis radius, and the periapsis velocity. Given the argument of periapsis, the orbital energy before and after perturbation can be derived according to the periapsis radius and the periapsis velocity. Simulations have been performed for orbits in the gravitational field of 216 Kleopatra. The numerical results are well consistent with analytic predictions.

## 1 Introduction

Asteroids are special bodies in the solar system that orbit the sun like planets but are much smaller in size. The shapes of asteroids are quite different from those of planets. Rather than spherical or quasi-spherical, asteroids are usually irregular in shape as compared to planets. Moreover, observations show that asteroids exhibit rapid rotation, with periods ranging from 2 hours to 20 hours[1]. The combination of these factors gives rise to a dynamic environment in the vicinity of an asteroid that is extremely complex, leading to abundant dynamic phenomena in this field.

Recently, orbital missions to asteroids, such as the Dawn mission and Hayabusa 2 mission, have begun to play an increasingly important role in deep space exploration[2, 3]. The dynamic environment surrounding asteroids is in some cases much more different from the types of orbital environment that have been studied in the past. Traditional studies of orbital motion have usually focused on the planetary case, around which there exists a central gravitational field. The equations of motion can be solved analytically in this case, and the corresponding orbital motion has been thoroughly studied[4]. However, asteroids constitute much more complicated cases in which traditional assumptions no longer apply, especially near the surface of asteroids. Considering the irregular shapes of asteroids, gravitational perturbations can be very large; thus, spacecraft would face great risks of impacting the surface when flying close to asteroids. To avoid these risks, further research is needed to estimate the orbital motion in the vicinity of asteroids.

One method of investigating the dynamic environment around asteroids is to use a triaxial ellipsoid to replace an asteroid, which can mainly describe the shape and size of an asteroid[5, 6, 7]. This gravitational model is simple and can be used to study the asteroid problem analytically. The drawback is that there is a great difference between the triaxial ellipsoid and real asteroids. Zhuravlev studied the stability of the libration points of a rotating triaxial ellipsoid[6, 7]. Using the Hamiltonian function and Lyapunov criterion, the stable and unstable regions can be obtained in the plane of parameters of the ellipsoid. Scheeres classified asteroids into two types according to the stability of equilibrium points. It has been proved that saddle equilibrium points are usually unstable and center equilibrium points can be either stable or unstable, depending on the characteristics of a given asteroid. Planar periodic orbits around Vesta and Eros have also been computed based on the triaxial ellipsoid model[5].

Expanding the gravitational potential into a harmonic series is another way to study orbital motion around asteroids. Due to the irregular shape of asteroids, the second degree and order gravity field terms are dominant in the harmonic series. Moreover, by using the harmonic series method, the equations of motion near asteroids can be derived analytically. Thus, the method provides a good approximation to the dynamic environment around asteroids. Some studies have been conducted using this method[8-14]. Antreasian et al. designed a low-attitude pass over asteroid 433 Eros, according to the analysis of the $C_{22}$ gravitational term[8]. By using a uniformly rotating second degree and order gravity field, equilibrium points and period orbits have been studied[9, 10]. The ejection and capture dynamics of asteroids have been studied by considering gravity potential of the second degree and order only. These effects appear to be determined by particle orbit and gravity coefficient $C_{22}$ in the case of retrograde motion[11]. Moreover, the $C_{22}$ gravitational term has been isolated to investigate the secular changes of orbital elements[12-14].

There are many types of orbits in the vicinity of asteroids. Previous studies have mostly concentrated on the impact of the dynamic environment of an asteroid on orbital motion, without actually addressing the structure of orbits. In this paper, we present a method for classifying the orbits near asteroids. A polyhedral approach was used to evaluate asteroid gravity fields[15]. The physical and geometrical characteristics of an asteroid can be obtained from Doppler-Radar images taken by ground-based observations. Therefore, a polyhedral asteroid model can be derived using a method proposed by Hudson. According to different structures of particle orbits near asteroids, which will undergo violent gravitational perturbation, orbits can be classified into 9 different categories. We used a second degree and order gravity field to study orbital characteristics analytically. When considering a direct orbit in the equatorial plane, the variation in the energy of the orbit is related to its argument of periapsis, periapsis radius, and velocity in periapsis. We present a method for analytically predicting orbit type by providing these three parameters for a certain asteroid, without computing the specific particle motion in the gravity field. Moreover, particle orbits in a polyhedron gravity field can be simulated by the forward and backward integration of the equations of motion at periapsis when the argument of periapsis, periapsis radius, and velocity at periapsis are given as the initial values. Thus, the validity of the predictions made by the analytical method can be examined. The results show that this method is valid in most cases, except for some critical situations, and will play an important role in establishing orbit types near asteroids for exploration tasks.

This paper is organized as follows: In Sect. 2, we introduce a polyhedral approach to estimate the gravitational field around asteroids and put forward the equations of motion for a small particle. In Sect. 3, we present a method for classifying orbits in the vicinity of an asteroid. In Sect. 4, we use harmonic expansion to derive the orbital potential and concentrate on the second degree and order gravitational field terms. A method for the prediction of orbital type is presented without numerical computations, which would prove quite helpful to the design of spacecraft orbits near asteroids. In Sect. 5, we use a model of asteroid 216 Kleopatra to determine orbital type and compare the results with the prediction results presented in Sect. 4. The results were observed to be consistent in most cases. And conclusions are finally drawn in Sect. 6.

**2 Gravitational Field**

To compute particle motion near an asteroid, a precise gravity field must be built outside the small body. In this study, we used a polyhedral approach to simulate the gravity field of asteroids and derived the equations of motion in a body-fixed, uniformly rotating frame. Given the initial particle conditions, the entire orbit can be obtained through numerical computation.

2.1 Polyhedral Method

The polyhedral method has been studied for decades. The history of this method is well explained in, in which a full description of modeling the gravity field of a small body using a constant-density polyhedron is presented[15].

By definition, the gravitational potential is expressed as follows:

$$U = G \iiint_M \frac{1}{r} dm. \tag{1}$$

An asteroid can be modeled as a constant-density polyhedron consisting of abundant planar faces. Thus, we can replace the gravity field of an asteroid with that of a polyhedron. Using a few mathematical operations, the potential can be expressed as a summation[15]:

$$U = \frac{1}{2} G\sigma \sum_{e \in edges} \mathbf{r}_e \cdot \mathbf{E}_e \cdot \mathbf{r}_e \cdot L_e - \frac{1}{2} G\sigma \sum_{f \in faces} \mathbf{r}_f \cdot \mathbf{F}_f \cdot \mathbf{r}_f \cdot \omega_f, \tag{2}$$

where $G=6.67\times10^{-11}$ $m^3kg^{-1}s^{-2}$ represents the gravitational constant, $\sigma$ is the density of the polyhedron, $\mathbf{r}_a$ ($a=e, f$) is a body-fixed vector from the field point to any point on an edge (corresponding to subscript $e$) or face (corresponding to subscript $f$), $L_e$ and $\omega_f$ are factors of integration that operate over the space between the field point and edges or faces, and $\mathbf{E}_e$ and $\mathbf{F}_f$ are dyads representing geometric parameters of edges and faces, which are defined in terms of face- and edge-normal vectors.

Thus, the polyhedron attraction and gravity gradient matrix can be derived as follows:

$$\nabla U = -G\sigma \sum_{e \in edges} \mathbf{E}_e \bullet \mathbf{r}_e \cdot L_e + G\sigma \sum_{f \in faces} \mathbf{F}_f \bullet \mathbf{r}_f \cdot \omega_f, \tag{3}$$

$$\nabla(\nabla U) = G\sigma \sum_{e \in edges} \mathbf{E}_e \cdot L_e - G\sigma \sum_{f \in faces} \mathbf{F}_f \cdot \omega_f. \tag{4}$$

Furthermore, the Laplacian of the polyhedron's potential can also be derived:

$$\nabla^2 U = -G\sigma \sum_{f \in faces} \omega_f, \tag{5}$$

It can be used to determine whether the field point is in the body of an asteroid or not, which is very useful in performing numerical computation. The polyhedron gravity field has sufficient precision to describe the dynamic environment, which is also valid near or on the surface of the polyhedron.

2.2 Equations of Motion

Here, we discuss a simple case: the asteroid is assumed to be a uniform rotator. To describe the orbits near an asteroid, the equations of motion are written in the body-fixed frame of the asteroid, that is, a uniformly rotating frame. The gravity field is time-variant in the inertial frame due to the rotation of the asteroid. However, in the body-fixed frame of the asteroid, the gravitational terms become time-invariant. They are only a function of the position of the field point. When we obtain the results in the body-fixed frame, the corresponding results in the inertial frame can be derived through a transformation matrix.

The origin of the body-fixed frame $O$ is defined to be at the center of mass. The $x$-axis lies along the smallest principal axis of the asteroid, and the $z$-axis lies along the largest principal axis. The $O$-$x$-$y$-$z$ coordinate frame is a right-handed coordinate system. Thus, the equations of motion in the body-fixed frame can be written as

$$\ddot{\mathbf{r}} + 2\boldsymbol{\Omega} \times \dot{\mathbf{r}} + \boldsymbol{\Omega} \times (\boldsymbol{\Omega} \times \mathbf{r}) = \nabla U(\mathbf{r}), \tag{6}$$

where $\mathbf{r}$ is a body-fixed vector, $|\boldsymbol{\Omega}| = \omega$ is the rotation velocity of the asteroid, and $U(\mathbf{r})$ is the gravitational potential. Assuming that the asteroid rotates around the largest principal axis, the direction of $\boldsymbol{\Omega}$ lies along the $z$-axis. The component form of the equations of motion can be expressed as

$$\ddot{x} - 2\omega\dot{y} = \omega^2 x + \frac{\partial U}{\partial x}, \tag{7}$$

$$\ddot{y} + 2\omega\dot{x} = \omega^2 y + \frac{\partial U}{\partial y}, \tag{8}$$

$$\ddot{z} = \frac{\partial U}{\partial z}. \tag{9}$$

Eq. (6) can be simplified by defining $V = 1/2\omega^2(x^2+y^2) + U$; thus,

$$\ddot{\mathbf{r}} + 2\Omega \times \dot{\mathbf{r}} = \nabla V. \tag{10}$$

An integral constant can be derived from Eq. (10).

$$J = \frac{1}{2}\dot{\mathbf{r}}\cdot\dot{\mathbf{r}} - \frac{1}{2}(\Omega \times \mathbf{r})\cdot(\Omega \times \mathbf{r}) - U = \frac{1}{2}\dot{\mathbf{r}}\cdot\dot{\mathbf{r}} - V, \tag{11}$$

where all the quantities are as previously defined. This integral is known as the Jacobi integral, which is conserved in the three-body problem. Once the position and velocity of the field point are given, the Jacobi integral will not change as the particle moves in the gravitational field.

**3 A Method for Classifying Orbits near Asteroids**

The dynamic environment around planets is usually described using central gravitational fields. Thus, the problem can be solved using a two-body model[4]. In this case, the orbital energy is an integral constant of the dynamic equations, and orbits can be classified into 3 categories: hyperbolic, parabolic, and elliptic, respectively, when the orbital energy is larger than zero, equal to zero, and less than zero. This represents the conic method of orbital classification, and the orbits are called Keplerian orbits. Furthermore, when we consider a more precise model of the planetary gravitational field, non-spherical gravitational perturbation must be involved in the equations of motion. Traditionally, the gravitational potential is expanded into a harmonic series and truncated at finite degree and order. Then, the variation in the orbit can be approximated using perturbation theory. For some major planets that are spherical in shape, the effect of non-spherical perturbation can be very minute. Orbital elements appear to vary only in the secular time period and are almost invariant over a short time period. In this case, the conic method of orbital classification still works well to describe orbits near celestial bodies. However, for asteroids, the situation is much different. Orbits near an asteroid are much more complicated compared with Keplerian orbits. The gravitational perturbation near an asteroid is so large that orbits sharply change their shape. For example, an orbit near an asteroid initiated as a hyperbola is very likely to change into an elliptic orbit after encountering the asteroid, and this type of orbit cannot be classified as a Keplerian orbit. Moreover, because the shapes of asteroids are usually irregular, for a trajectory inside a circumscribing sphere, it becomes more difficult to predict whether the trajectory intersects the surface of an asteroid. In fact, there can even be a period orbit inside the circumscribing sphere. However, in the case of a major planet, this prediction can be very simple to make by comparing the periapsis radius of the trajectory with the radius of the planet. Therefore, the conic method of orbital classification does not work in the asteroid case.

For orbits near an asteroid, the periapsis is the closest point to the center of mass, which means that the gravitational perturbation can be very strong and the energy of the orbit will change sharply. Similar to the case of Keplerian orbits, the orbital energy has a direct influence on the orbit shape.

This paper presents a method for classifying orbits near asteroids. According to the variation in their shapes after passing the periapsis, the orbits near an asteroid are classified into 9 categories. Table 1 shows the different orbital types. Before flying by the periapsis, an orbit can surround an asteroid, fly from infinity, or fly from the surface. Analogously, after flying by the periapsis, an orbit surround an asteroid, fly to infinity, or encounter the surface. By combining these possible situations, 9 different orbit types can be determined.

In Table 1, "surface" represents the type of orbits touching the surface of an asteroid; "infinity" represents the type of orbits flying from or to infinity; "surrounding" represents the type of orbits surrounding an asteroid. Because the dynamics of an asteroid is much more complicated than that of an planetary, it is difficult to determine whether an orbit will surround an asteroid permanently. Therefore, the "surrounding" orbits presented in this paper are all temporary. "Surrounding-to-surrounding" orbits indicate those ones that surround an asteroid before flying by the periapsis and also surround it after the flyby, and other types of orbits are defined similarly.

**Table 1** 9 Types of Orbits near Asteroids

| Orbital Types | | |
|---|---|---|
| Surrounding-to-surrounding | Surrounding-to-surface | Surrounding-to-infinity |
| Infinity-to-surrounding | Infinity-to-surface | Infinity-to-infinity |

| Surface-to-surrounding | Surface-to-surface | Surface-to-infinity |
|---|---|---|

The orbital energy varies significantly during the passage of the periapsis due to the great perturbation of the gravity field. The amount by which the energy changes varies greatly from orbit to orbit. "Surrounding-to-surrounding" orbits experience smaller changes in energy so that the orbits remain (temporarily) around the asteroid after the passage of the periapsis. However, for "infinity-to-surrounding" orbits, the change in energy is larger. Before the flyby of the periapsis, the orbits come from infinity and have positive orbital energy. Then, the orbital energy becomes negative after the passage of the periapsis. This means that the asteroid captures (temporarily) the particle from infinity.

The study of orbital classification near asteroids can be very useful in mission design. Researchers usually use a natural orbit as a nominal trajectory in the exploration of asteroids. If we choose the "infinity-to-surrounding" orbit as a nominal trajectory when designing an orbit near an asteroid, the "capture" characteristics can be used to slow a spacecraft down, which will dramatically reduce fuel consumption. After an orbit transitions to a surrounding orbit, a control procedure must be applied to make the orbit stable.

## 4 Analyses of Orbital Types

As mentioned above, we focused on the variation in orbital energy when classifying orbits. Thus, the perturbation equations of energy are set up at the periapsis point to theoretically analyze orbital types near an asteroid. We use a harmonic series method to expand the gravitational field of asteroids and investigate the influence of the second degree and order gravitational terms on the orbital energy.

The high degree and order harmonics of the gravitational field have a great impact on orbits near asteroids. Scheeres and Hu conducted much research in this vein. Based on their studies, we can derive the equations describing the variation in the energy of an orbit near asteroids by using the parameters of the periapsis[5, 9-14].

In classical orbital theory, the Keplerian energy of an orbit can be expressed as

$$C_2 = \frac{1}{2}\mathbf{v}_I \cdot \mathbf{v}_I - \frac{\mu}{|\mathbf{r}|}. \tag{12}$$

Here, $\mu$ represents the gravitational parameter of the center body, where $\mu = Gm$, $\mathbf{v}_I$ is the velocity vector in the inertial coordinate system, and $\mathbf{r}$ is the position vector of the particle. The velocity vector $\mathbf{v}_I$ can be expressed in the body-fixed system as

$$\mathbf{v}_I = \dot{\mathbf{r}} + \Omega \times \mathbf{r}. \tag{13}$$

The potential of the asteroid is written as

$$U(\mathbf{r}) = \frac{\mu}{|\mathbf{r}|} + \sum_i U_i(\mathbf{r}), \tag{14}$$

where $U_i(\mathbf{r})$ terms are the higher degree and order harmonics of the gravitational field, i=2,3,4... Substituting Eqs. (13) and (14) into Eq. (12) yields

$$C_2 = \sum_i U_i(\mathbf{r}) - U + \frac{1}{2}(\dot{\mathbf{r}} + \Omega_T \times \mathbf{r}) \cdot (\dot{\mathbf{r}} + \Omega_T \times \mathbf{r}). \tag{15}$$

Using Eq. (11), the above equation can be expressed as

$$C_2 = \sum_i U_i(\mathbf{r}) + (\Omega \times \mathbf{r}) \cdot (\dot{\mathbf{r}} + \Omega \times \mathbf{r}) + J, \tag{16}$$

where $J$ is the Jacobi integral as defined above.

For the purpose of determining the variation of orbital energy, we take the time differential of both sides of Eq. (16) in the body-fixed system, and obtain

$$\dot{C}_2 = \sum_i \nabla U_i(\mathbf{r}) \cdot \dot{\mathbf{r}} + (\ddot{\mathbf{r}} + \Omega \times \mathbf{r}) \cdot (\Omega \times \mathbf{r}). \tag{17}$$

The above equation can be simplified by using the equations of motion.

$$\dot{C}_2 = \sum_i \nabla U_i(\mathbf{r}) \cdot (\dot{\mathbf{r}} + \Omega \times \mathbf{r}) = \sum_i \nabla U_i(\mathbf{r}) \cdot \mathbf{v}_I = \sum_i U_i'(\mathbf{r}), \tag{18}$$

where the $U_i'$ terms correspond to the time derivatives of the higher degree and order harmonics in the body-fixed coordinate system.

Among all the higher degree and order harmonics, the second degree and order gravitational terms appear to dominate[11]. Setting $i=2$, the higher degree and order potential can be written as

$$U_2 = \frac{\mu}{r^3}\left[\frac{r_o^2 C_{20}}{2}(3\sin^2\alpha - 1) + 3r_o^2 C_{22}(1 - \sin^2\alpha)\cos(2\lambda)\right], \tag{19}$$

where $r_o$ is a normalized radius, whose value is arbitrarily determined. $\lambda$ and $\alpha$ are the longitude and latitude, respectively, of a testing point in the body-fixed system. $C_{20}$ and $C_{22}$ are the coefficients of the second degree and order harmonic terms.

Here, we investigate a simple case. Assuming that an orbit is fixed in the equatorial plane, we obtain $\alpha=0$ and $\lambda=\nu+f-\omega t$, where $\nu$ is the argument of periapsis. With this assumption and using Eq. (18), we obtain

$$\dot{C}_2 \approx \frac{9\mu e r_o^2 C_{22}\dot{f}}{a(1-e^2)r^2}\sin 2\nu \sin f \sin 2(f-\omega t) - \frac{6\mu r_o^2 C_{22}\dot{f}}{r^3}\sin 2\nu \cos 2(f-\omega t). \tag{20}$$

It is clear that only the $C_{22}$ term works. The orbital elements are time-variant during motion. The average values during the passage of the periapsis can be approximated by the orbital elements of the periapsis. Then, Eq. (20) is integrated in the vicinity of the periapsis.

$$\Delta C_{2,p} \approx -\frac{3\mu r_o^2 C_{22}\dot{f}_p}{r_p^3}\sin 2\nu\left[\frac{2\sin 2(\omega - \dot{f}_p)T}{\omega - \dot{f}_p}\right.$$
$$\left. + \frac{3e}{1+e}\left\{\frac{\sin(2\omega - 3\dot{f}_p)T}{2\omega - 3\dot{f}_p} - \frac{\sin(2\omega - \dot{f}_p)T}{2\omega - \dot{f}_p}\right\}\right], \tag{21}$$

where $2T$ is the integral time, from $-T$ to $+T$. Considering the acting time of the gravitational perturbation, we generally set $T=\pi/4\omega$, which corresponds to a eighth of the rotation period. $\dot{f}_p$ and $r_p$ are the angular velocity of true anomaly and radius, respectively. We also have the formula

$$\dot{f}_p = \sqrt{\mu(1+e)/r_p^3}. \tag{22}$$

The subscript $p$ indicates periapsis. Here, we obtain the variation in orbital energy. To investigate the change in orbital shape, the Keplerian energies before and after the passage of the periapsis are derived.

Here, we further assume that the orbital energy varies uniformly and continuously, with the symmetrical center located in the periapsis. Thus, before and after the particle goes through the periapsis, the variations in energy are equal in value: both are half of the total variation. We use $C_{2,p}$ to represent the Keplerian energy at the periapsis.

$$F_1 = C_{2,p} + 0.5\Delta C_{2,p}, \tag{23}$$

$$F_2 = C_{2,p} - 0.5\Delta C_{2,p}, \tag{24}$$

where $F_1$ is the Keplerian energy after the perturbation during the passage of the periapsis and $F_2$ is the Keplerian energy before the perturbation. The variation in energy is illustrated in Fig. 1.

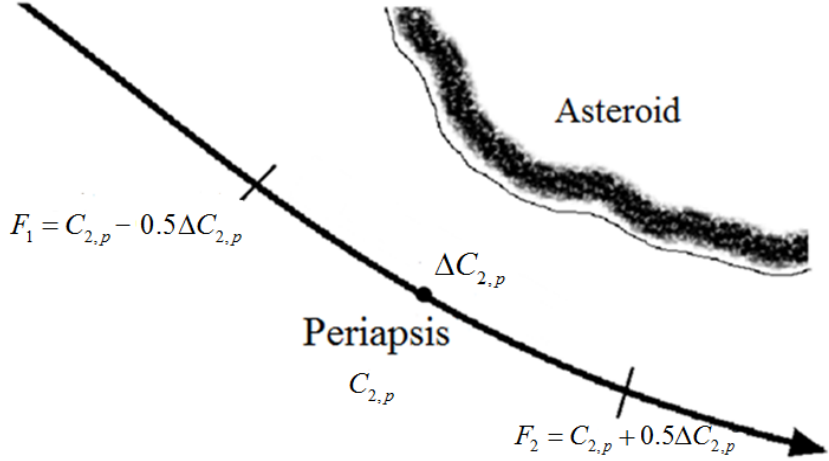

**Fig. 1** The conditions of variation in orbital energy. $F_1$ represents the Keplerian energy after the perturbation during the passage of the periapsis, and $F_2$ represents the Keplerian energy before the perturbation; $\Delta C_{2,p}$ represents the variation in energy during the passage of the periapsis.

By substituting Eqs. (12) and (21) into Eqs. (23) and (24), the following equations are obtained.

$$F_1 = -\frac{\mu(1-e)}{2r_p} - \frac{3\mu r_o^2 C_{22} \dot{f}_p \sin 2\nu}{2r_p^3}\left[\frac{2\sin 2(\omega - \dot{f}_p)T}{\omega - \dot{f}_p} + \frac{3e}{1+e}\left\{\frac{\sin(2\omega - 3\dot{f}_p)T}{2\omega - 3\dot{f}_p} - \frac{\sin(2\omega - \dot{f}_p)T}{2\omega - \dot{f}_p}\right\}\right],$$
$$\tag{25}$$

$$F_2 = -\frac{\mu(1-e)}{2r_p} + \frac{3\mu r_o^2 C_{22} \dot{f}_p \sin 2\nu}{2r_p^3}\left[\frac{2\sin 2(\omega - \dot{f}_p)T}{\omega - \dot{f}_p} + \frac{3e}{1+e}\left\{\frac{\sin(2\omega - 3\dot{f}_p)T}{2\omega - 3\dot{f}_p} - \frac{\sin(2\omega - \dot{f}_p)T}{2\omega - \dot{f}_p}\right\}\right],$$
$$\tag{26}$$

where all the variables are as previously defined. In the expression

$$v_p = \sqrt{\mu(1+e)/r_p}, \tag{27}$$

the values of $F_1$ and $F_2$ are dependent on the argument of periapsis $\nu$, periapsis radius $r_p$, and velocity at periapsis $v_p$, i.e., $F_1 = F_1(r_p, v_p, \nu)$, $F_2 = F_2(r_p, v_p, \nu)$. When $\nu$, $r_p$ and $v_p$ are given, we can predict the type of orbit according to the values of $F_1$ and $F_2$.

**5 Numerical Simulations**

The polyhedral model and physical parameters of asteroid 216 Kleopatra have already been obtained from ground-based observations. Numerical computations are conducted in the gravitational field of the polyhedral model.

5.1 Physical Parameters of Asteroid 216 Kleopatra

Asteroid Kleopatra was first discovered by Johann Palisa in 1880. It is one of the largest main-belt asteroids. Because of its rapid variation in light curve shape, Kleopatra has drawn widespread attention among researchers. Marchis et al. found that there are two natural satellites surrounding Kleopatra. Additionally, there are many researchers who are currently studying the asteroid's shape and physical characteristics based on ground-based observations. Ostro and Hudson generated a polyhedral model of Kleopatra[16]. Descamps and Marchis provided the physical parameters of 216 Kleopatra[17]. Yang and Hexi investigated the orbital dynamics near 216 Kleopatra[18].

**Table 2** Physical Parameters of 216 Kleopatra

| Mass/kg | Size/km | Average density/g/cm$^3$ | Rotation period/h |
|---|---|---|---|
| $(4.64\pm0.02)\times10^{18}$ | 217×94×81 | 3.6±0.4 | 5.385 |

The polyhedral model of 216 Kleopatra consists of 2048 vertexes and 4092 faces, and it is assumed to have constant density. Using the polyhedron method in the body-fixed coordinate system as defined previously, the three principle moments of inertia are obtained. The results are shown in Table 3.

**Table 3** Moments of Principle Axes of Kleopatra

|  | Moments of inertia/kg·km$^2$ |
|---|---|
| $I_x$ | 1.677128779395×10$^{21}$ |
| $I_y$ | 1.144620100312×10$^{22}$ |
| $I_z$ | $1.153288976690 \times 10^{22}$ |

When the potential of the gravitational field is expanded into a harmonic series, the coefficients of the second degree and order terms are related to the principle moments of inertia as follows:

$$C_{20} = \frac{1}{mr_o^2}\left[\frac{1}{2}(I_x + I_y) - I_z\right], \quad (28)$$

$$C_{22} = \frac{1}{4mr_o^2}\left[I_y - I_x\right], \quad (29)$$

where all the quantities are as previously defined. Setting $r_o=1$ km, we obtain

$$C_{20} = -1948.0292 \quad (30)$$

$$C_{22} = 957.02962 \quad (31)$$

5.2 Orbital Type Prediction

By substituting the corresponding quantities for asteroid Kleopatra into Eq. (25) and Eq. (26), the effect of $r_p$ and $v_p$ on $F_1$ and $F_2$ are investigated by fixing the argument of periapsis $v$. Fig. 2 and Fig. 3 show the contour plots of $F_1$ and $F_2$ when $\sin 2v=1$ (i.e., $v=\pi/4+k\pi$) are taken in the $(r_p, v_p)$ plane. Note that because $\Delta C_{2,p}$ is an odd function of the argument of periapsis $v$, the values of $F_1(r_p, v_p)$ and $F_2(r_p, v_p)$ are simply interchanged when $\sin 2v=-1$ (i.e., $v=3\pi/4+k\pi$). That is, when $\sin 2v=-1$, the contour plot of $F_1(r_p, v_p)$ corresponds to that shown in Fig. 3 and the plot of $F_2(r_p, v_p)$ corresponds to that shown in Fig. 2.

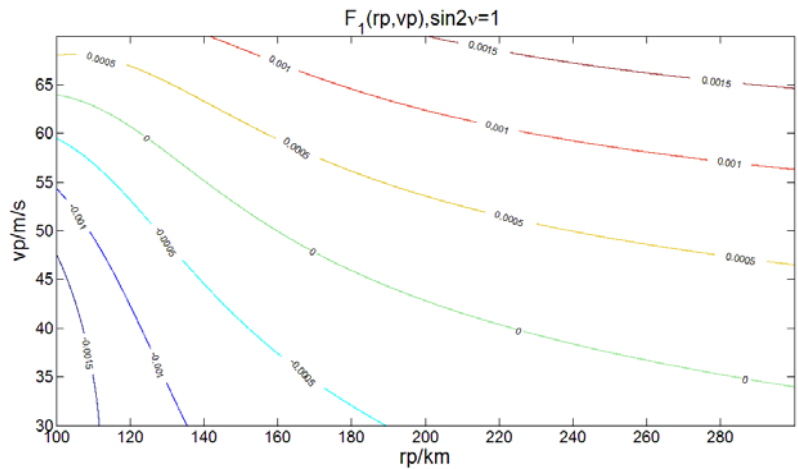

**Fig. 2** Contour plot of $F_1$ when $\sin 2v=1$ for Kleopatra. From bottom left to upper right, the value

of $F_1$ increases from less than zero to greater than zero. When sin2$v$=−1, the contour plot corresponds to that of $F_2$.

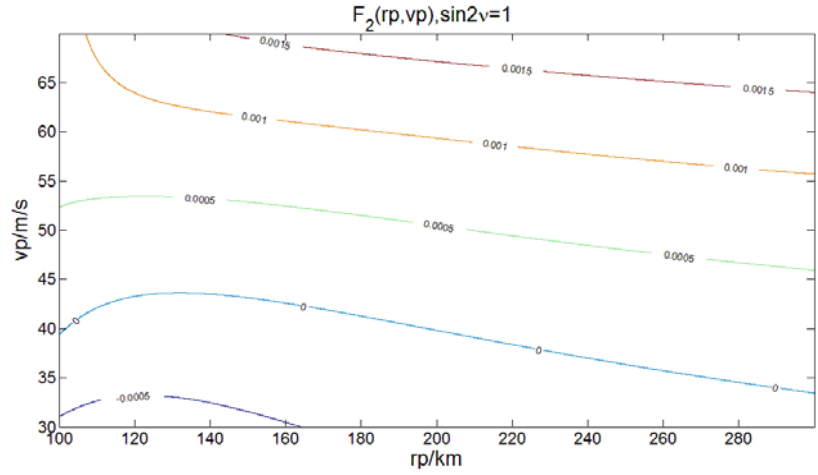

**Fig. 3** Contour plot of $F_2$ when sin2$v$=1 for Kleopatra. From bottom left to upper right, the value of $F_1$ increases from less than zero to greater than zero. When sin2$v$=−1, the contour plot corresponds to that of $F_1$.

Fig. 2 and Fig. 3 show that above the curve of $F_1$=0 ($F_2$=0) the value of $F_1$($F_2$) is positive and below the curve of $F_1$=0 ($F_2$=0) the value is negative. For different values of $v$, the curves of $F_1$=0 and $F_2$=0 are located in different positions of the figure. With the decrease in the value of sin2$v$, the curve of $F_1$=0 moves downward and the curve of $F_2$=0 moves upward in the ($r_p$,$v_p$) plane. When sin2$v$=−1, the curve of $F_1$=0 reaches the position that the curve of $F_2$=0 occupies for sin2$v$=1. A similar situation occurs for the curve of $F_2$=0. In the ($r_p$,$v_p$) plane, the curve $F_1$=0 represents the critical state of orbital energy after the passage of the periapsis, indicating a parabolic trajectory. It also represents the intermediate state among orbital styles. A positive value of $F_1$ indicates that the orbital energy is greater than zero after the passage of the periapsis. Afterward, the particle escapes the gravitational field and flies to infinity. A negative value of $F_1$ indicates that the orbital energy is less than zero after the passage of the periapsis. Thus, it cannot break away from the gravitational field and stays in the vicinity of the center body. Similarly, the curve of $F_2$=0 represents the critical state of orbital energy before the passage of the periapsis. If the value of $F_2$ is greater than zero, the particle initially flies from infinity. Otherwise, it initially stays in the vicinity of the small body.

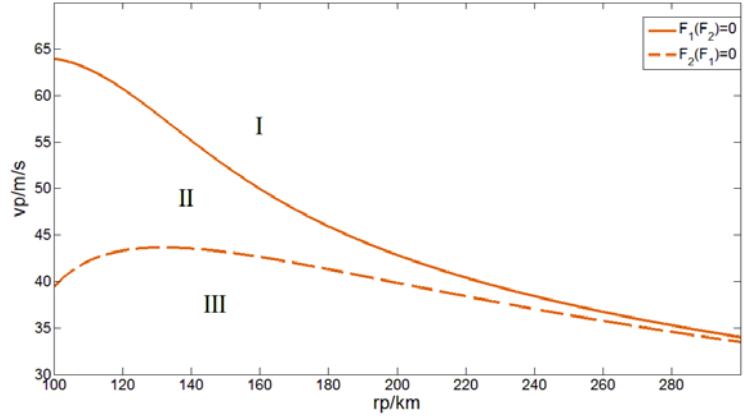

**Fig. 4** The critical conditions of $F_1$=0 and $F_2$=0. When $\sin 2\nu$=1, the solid line is $F_1$=0 and the dashed line is $F_2$=0. When $\sin 2\nu$=−1, the dashed line is $F_1$=0 and the solid line is $F_2$=0.

Fig. 4 shows the critical state of curves $F_1(r_p,v_p)$=0 and $F_2(r_p,v_p)$=0. When $\sin 2\nu$=1, the upper curve indicates $F_1(r_p,v_p)$=0 and the lower curve indicates $F_2(r_p,v_p)$=0. When $\sin 2\nu$=−1, the opposite is true. By using Fig. 4, we can predict the orbital type about a point in the $(r_p,v_p)$ plane. Based on the relative position between curve $F_2(r_p,v_p)$=0 and a point in the $(r_p,v_p)$ plane, the energy before the passage of the periapsis passage can be determined. Based on the relative position between the point and curve $F_1(r_p,v_p)$=0, the energy after the passage of the periapsis can also be determined. By combining these two results, the orbital types can be predicted. The $(r_p,v_p)$ plane is separated into three parts by curves $F_1(r_p,v_p)$=0 and $F_2(r_p,v_p)$=0. We mark them as region I~III.

**Table 4** Possible Regions for Each Type of Orbit with Different Arguments of Periapsis

|  | $\sin 2\nu$=1 | $\sin 2\nu$=−1 |
| --- | --- | --- |
| Surrounding-to-surrounding | III | III |
| Surrounding-to-surface | III | (II),III |
| Surrounding-to-infinity | —— | II |
| Infinity-to-infinity | I | I |
| Infinity-to-surrounding | II | —— |
| Infinity-to-surface | (I),II | (I) |
| Surface-to-surface | (I),(II),III | (I),(II),III |
| Surface-to-surrounding | (II),III | III |
| Surface-to-infinity | (I) | (I),II |

When $\sin 2\nu=1$, region I and region II are on the upper side of curve $F_2(r_p,v_p)=0$. Therefore, the energy before the passage of the periapsis is greater than zero, which corresponds to a hyperbolic trajectory. Region III is on the lower side of curve $F_2(r_p,v_p)=0$. Therefore, the energy before the passage of the periapsis is less than zero, which means that the orbit is in the vicinity of the asteroid. Region I is on the upper side of curve $F_1(r_p,v_p)=0$. Therefore, the energy after the perturbation of the passage of the periapsis is greater than zero, indicating a hyperbolic trajectory. Region II and region III are both on the lower side of curve $F_1(r_p,v_p)=0$, which means that the orbital energy after the perturbation of the passage of the periapsis is negative. A similar analysis is performed when $\sin 2\nu=-1$. For different values of $\nu$, the size of each region will be correspondingly different. The area of region II is the largest when $\sin 2\nu=\pm 1$. However, when $\sin 2\nu=0$, curves $F_1(r_p,v_p)=0$ and $F_2(r_p,v_p)=0$ overlap and region II disappears. Each point in the $(r_p,v_p)$ plane corresponds to a certain orbit. Thus, the conditions illustrated in Fig. 4 are closely related to the orbital type. For example, for the "infinity-to-surrounding" orbital type, the necessary condition is that the orbital energy is positive before the perturbation in the passage of the periapsis ($F_2(r_p,v_p)>0$) and negative after the passage of the periapsis ($F_1(r_p,v_p)<0$). When $\sin 2\nu=1$, the orbit can take place in region II. The other 8 orbital types can be analyzed similarly.

Table 4 presents the possible regions where each orbital type may take place. We can see that there are two orbital types, "infinity-to-surrounding" and "surrounding-to-infinity", that take place in only one region, respectively. These special orbits are related to the capture and ejection dynamics of asteroids. Scheeres has conducted intensive studies on these effects[11].

It is important to note the orbital energy of the condition "surface". Generally, the condition "surface" corresponds to any range of energy because it simply indicates the positional relationship between an orbit and the surface, not an energy condition. If an orbital type is from "surface" to another condition, it may fly from the surface of an asteroid at any velocity. This means that the energy can be either positive (if the launch velocity is high enough) or negative (if the launch velocity is low). If an orbital type is from some other condition to "surface", it may be the case in which the orbital altitude descends after the perturbation during the passage of the periapsis so that it touches the surface of an asteroid, and the energy decreases to a value below zero, or the orbital energy increases to a value greater than zero. However, the periapsis radius is quite small and thus very close to the asteroid. Meanwhile, the shapes of asteroids are very irregular. Therefore, the orbit still touches the surface of the asteroid while the orbital energy is positive. There are some situations shown in brackets in Table 4 that all correspond to the "surface" condition. Only if the orbit is very close to the asteroid or if other special conditions apply, which is rare in most cases, can these situations occur. Here, we only note the possibility that these situations occur and do not analyze them in the subsequent discussion.

5.3 Computing Orbital Types

We make many assumptions when deriving the variation in energy during the passage of the periapsis. Thus, the prediction results can be different from those observed under actual conditions. In this section, we choose some points from each region in the $(r_p, v_p)$ plane and run some simulations in the polyhedron gravitational field of asteroid 216 Kleopatra. Thus, the calculation results can be obtained and compared with those shown in Table 4 to verify our prediction method.

Three points in the $(r_p, v_p)$ plane, A (160 km, 60 m/s), B (160 km, 47 m/s), and C (160 km/s, 34 m/s), are taken as the initial conditions; these points are located in regions I, II, and III, respectively. For $\sin 2\nu = 1$ and $\sin 2\nu = -1$, the orbital types can be computed in the polyhedral gravitational field.

a. A(160 km, 60 m/s), $\sin 2\nu = 1$.

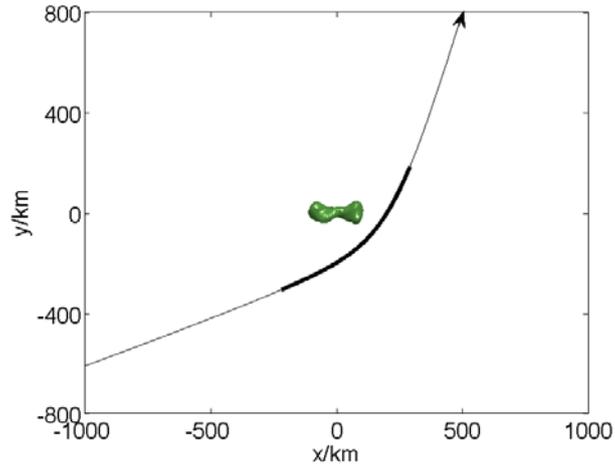

(a) Orbital type under initial condition a.

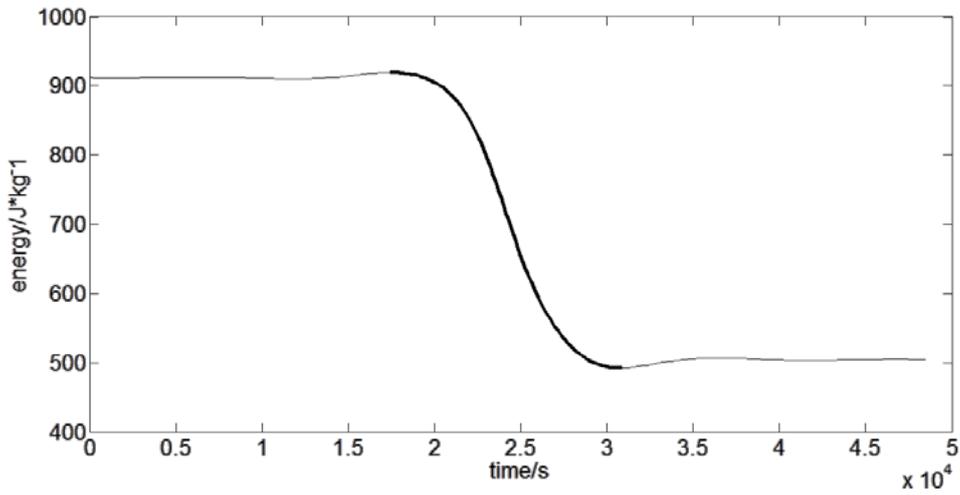

(b) The variation in orbital energy under condition a.

**Fig. 5** Computational results with initial condition a. Noting that the asteroid rotates in the inertial system, we only illustrate the initial position of the asteroid. The bold section corresponds to the passage of the periapsis, which features great perturbation. The same notion applies to the following figures.

Fig. 5 indicates that the orbital type is "infinity-to-infinity". The trajectory first goes from infinity with positive orbital energy 910 J/kg. When going through the periapsis, the energy decreases to 500 J/kg, which is still greater than zero. Thus, it can escape from the asteroid and ultimately go to infinity. The perturbation time (bold) lasts for approximately $1\times10^4$ s.

b. A(160 km, 60 m/s), $\sin 2\nu = -1$.

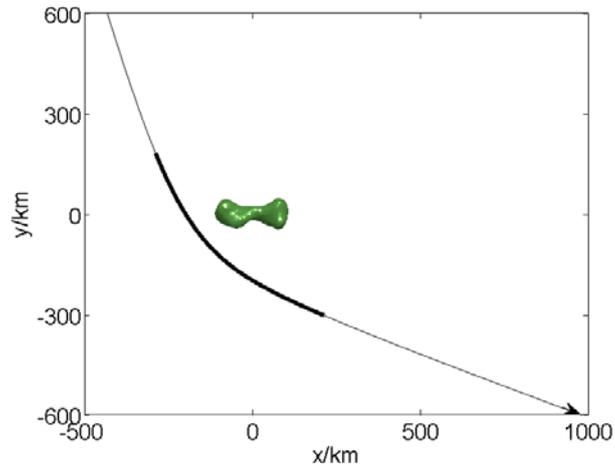

(a) Orbital type under condition b.

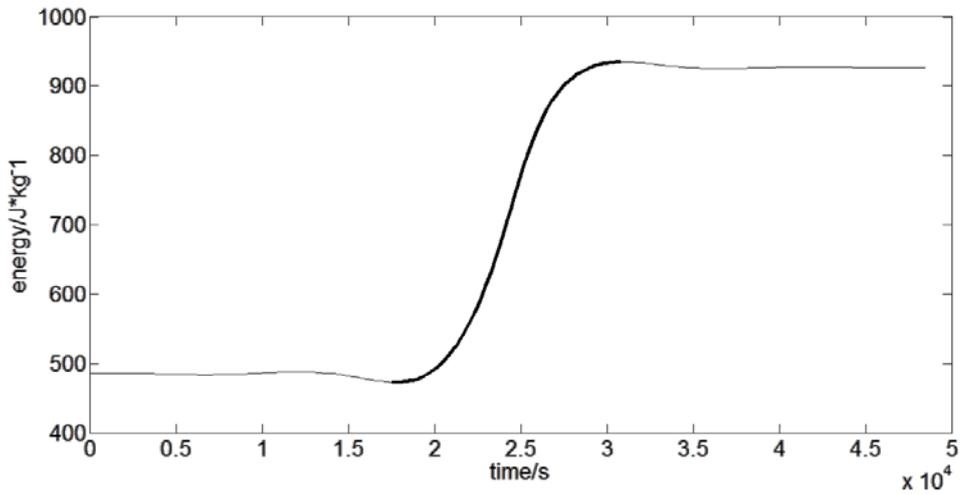

(b) The variation in orbital energy under condition b.

**Fig. 6** Computational results with initial condition b.

Fig. 6 indicates that the orbital type is "infinity-to-infinity". The trajectory first goes from infinity with positive orbital energy 490 J/kg. When going through the periapsis, the energy increases to 910 J/kg, which is also greater than zero. Thus, it can escape from the asteroid and go to infinity. The perturbation time (bold) lasts for approximately $1\times10^4$ s.

c. B(160 km, 47 m/s), $\sin 2\nu = 1$.

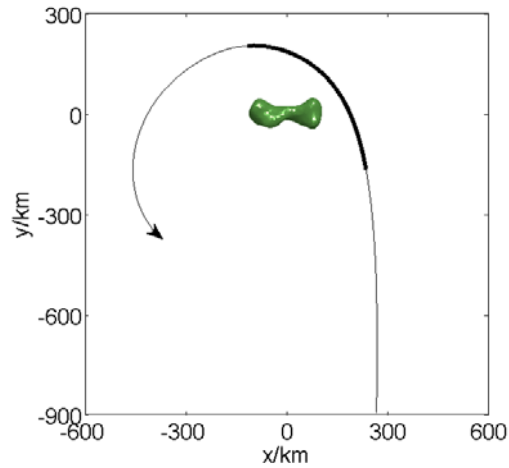

(a) Orbital type under condition c.

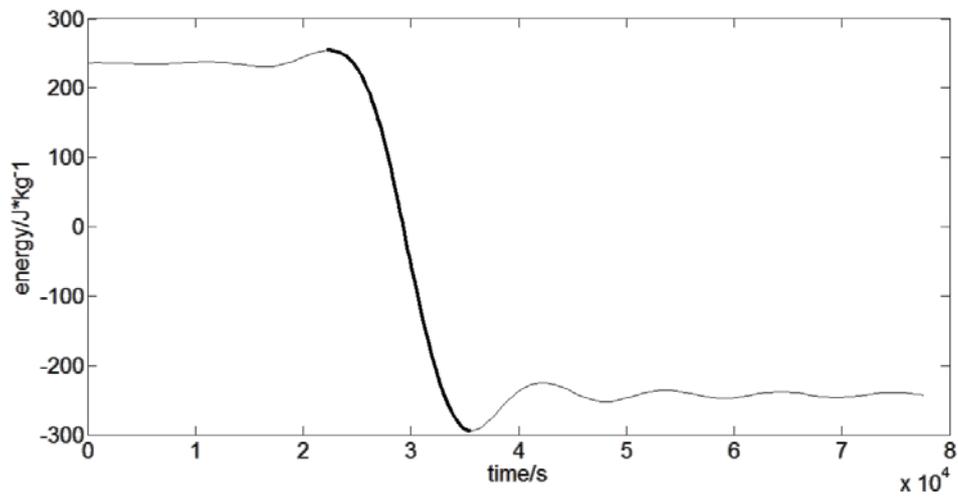

(b) The variation in orbital energy under condition c.

**Fig. 7** Computational results with initial condition c.

Fig. 7 indicates that the orbital type is "infinity-to-surrounding". The trajectory first goes from infinity with positive orbital energy 230 J/kg. When going through the periapsis, the energy decreases to −290 J/kg, which is less than zero. Thus, it turns into a surrounding (temporarily) orbit. The perturbation time (bold) lasts for nearly $1\times10^4$ s.

d. B(160 km, 47 m/s), $\sin 2\nu = -1$.

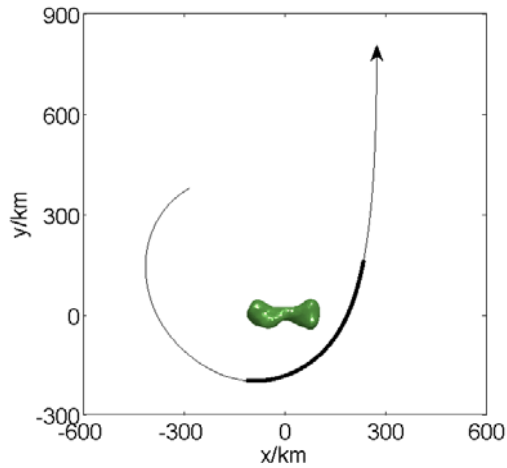

(a) Orbital type under condition d.

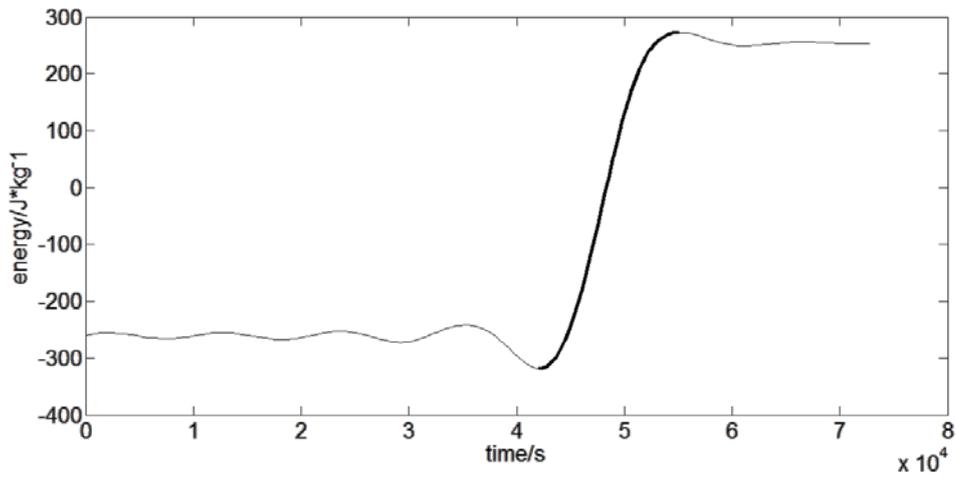

(b) The variation in orbital energy under condition d.

**Fig. 8** Computational results with initial condition d.

Fig. 8 indicates that the orbital type is "surrounding-to-infinity". The trajectory is a surrounding orbit initially. The orbital energy is −270 J/kg. When going through the periapsis, the energy increases to 270 J/kg, which is greater than zero. Thus, it escapes from the asteroid and goes to infinity. The perturbation time (bold) lasts for $1\times10^4$ s.

e. C(160 km/s,34 m/s), $\sin2\nu=1$.

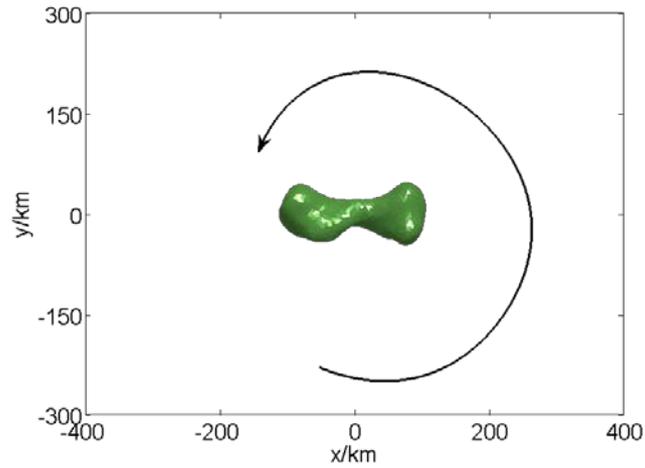

(a) Orbital type under condition e.

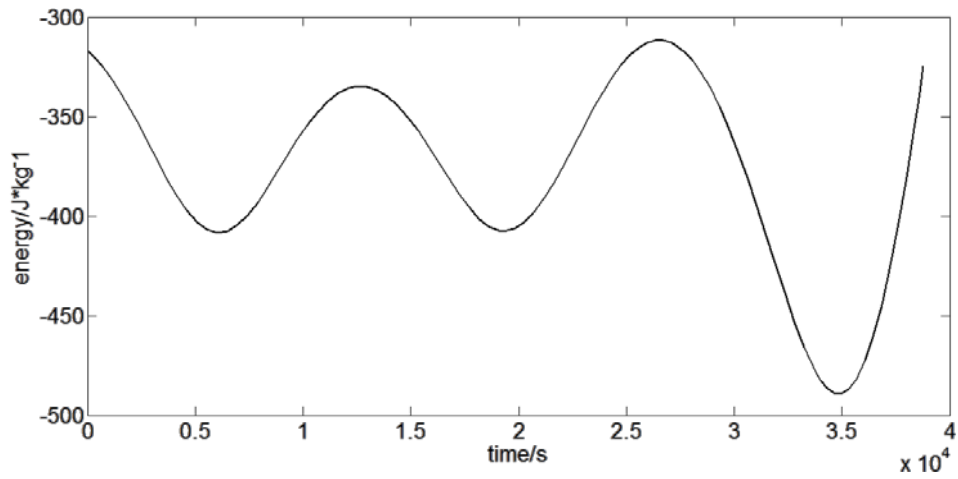

(b) The variation in orbital energy under condition e.

**Fig. 9** Computational results with initial condition e.

Fig. 9 indicates that the orbital type is "surrounding-to-surrounding". The orbital energy is less than zero initially and fluctuates over the simulation time. Passage through the periapsis is not obvious.

f. C(160 km/s,34 m/s), sin2$\nu$=−1.

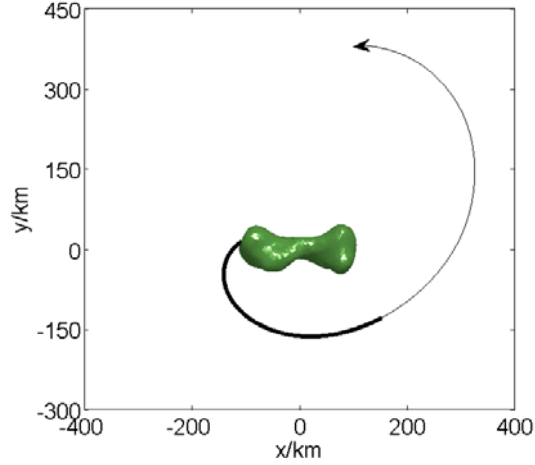

(a) Orbital type under condition f.

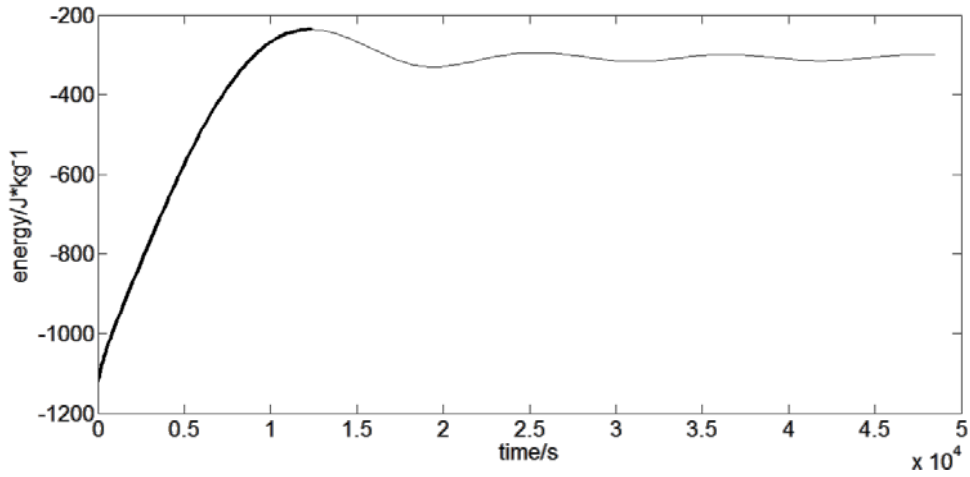

(b) The variation in orbital energy under condition f.

**Fig. 10** Computational results with initial condition f.

Figure 10 indicates that the orbital type is "surface-to-surrounding". The trajectory first launches from the asteroid with orbital energy $-1100$ J/kg. When going through the periapsis, the energy increases to $-300$ J/kg. The orbital energy is still less than zero, and the trajectory surrounds the asteroid. The perturbation time (bold) is nearly $1\times10^4$ s.

The simulation results are presented in Table 5. Comparing these results with those shown in Table 4, we can see that the predictions agree with the simulation results. When $\sin 2\nu=1$ is changed to $\sin 2\nu=-1$, the locations of curves $F_1(r_p,v_p)=0$ and $F_2(r_p,v_p)=0$ will interchange. This means that for the same parameters $(r_p,v_p)$, the initial state and final state will interchange when the $\sin 2\nu$ term changes its sign. This is a type of symmetry for orbital structures. When focus is laid on condition a and condition b, this symmetry can be noticed. Under condition a, the initial orbital energy is 910 J/kg and the final orbital energy is 500 J/kg. The perturbation time is $1\times10^4$ s. Under condition b, the initial orbital energy is 490 J/kg and the final orbital energy is 910 J/kg. The perturbation time is $1\times10^4$ s. Because the shape of Kleopatra is not strictly symmetrical, the energy is also not strictly symmetrical. When an orbit is of the "surface" type, symmetry does not generally exist. The shapes of asteroids are usually irregular, and predicting orbits of the "surface" type is difficult. This difficulty varies from case to case, and the results depend on the particular situation being addressed. Condition e and f are examples of the cases without symmetry.

**Table 5** Computational Results of Orbital Types in the Gravitational Field of Kleopatra

|  | $\sin2\nu=1$ | $\sin2\nu=-1$ |
|---|---|---|
| A(160 km/s,60 m/s) | infinity-to-infinity | infinity-to-infinity |
| B(160 km/s,47 m/s) | infinity-to-surrounding | surrounding-to-infinity |
| C(160 km/s,34 m/s) | surrounding-to-surrounding | surface-to-surrounding |

## 6 Conclusions

A new method for classifying orbits is presented to analyze the dynamic environment near an asteroid. According to the variation in orbital shape after the passage of the periapsis, orbits are classified into 9 types: "surrounding-to-surrounding", "surrounding-to-infinity", "surrounding-to-surface", "infinity-to-surrounding", "infinity-to-surface", "infinity-to-infinity", "surface-to-surface", "surface-to-surrounding", and "surface-to-infinity". For a certain asteroid, the orbital type can be predicted when the parameters of periapsis (argument of periapsis, periapsis radius, and velocity at periapsis) are given. We compute the orbital types using a polyhedral model of asteroid 216 Kleopatra. The results of the simulation show that the prediction of orbital type is correct in some aspects. But in some special cases, for example, the orbit is very near the surface of an asteroid, this prediction may be unusable. This method of orbital classification and prediction is also valid for other asteroid models.